\documentstyle[prd,aps,twocolumn,epsf,epsfig]{revtex}

\def\br{\begin{eqnarray}}
\def\er{\end{eqnarray}}
\def\be{\begin{equation}}
\def\ee{\end{equation}}
\def\a{\alpha}
\def\b{\beta}

\def\d{\delta}

\def\L{\Lambda}
\def\m{\mu}

\def\ad {{\dot\alpha}}
\def\md {{\dot\mu}}
\def\bd {{\dot\beta}}

\def\N {{\nabla}}
\def\Nb {{\bar\nabla}}

\def\({\left(}
\def\){\right)}
\def\<{\left\langle}
\def\>{\right\rangle}

\begin{document}
\twocolumn[\hsize\textwidth\columnwidth\hsize\csname 
@twocolumnfalse\endcsname                            
\title{Consistency of Superspace Low Energy Equations of Motion of 4D Type II Superstring with Type II Sigma Model at Tree-Level}
\author{D. L. Nedel}
\address{Instituto de F\'{\i}sica Te\'orica, UNESP,
Rua Pamplona 145, 01405-900, S\~ao Paulo, SP, Brazil}
\date{\today}

\maketitle

\begin{abstract}
We derive the torsion constraints and show the consistency of equations of motion of four-dimensional Type II supergravity in superspace, with Type II sigma model. This is achieved by coupling the
four-dimensional compactified Type II Berkovits' superstring to an N=2 curved
background and requiring that the sigma-model has superconformal invariance at tree-level. We compute this in a manifestly 4D N=2 supersymmetric way. The
constraints break the target conformal and SU(2) invariances and the dilaton
will be a conformal, $SU(2)\times U(1)$ compensator. 
For Type II superstring in four dimensions, worldsheet
supersymmetry requires two different compensators. One type is described by
chiral and anti-chiral superfields. This compensator can be
identified with a vector multiplet. The other Type II compensator is
described by twist-chiral and twist-anti-chiral superfields and can be identified with a tensor hypermultiplet. Also, the superconformal invariance at tree-level selects a particular gauge, where the matter is fixed, but not the compensators. After imposing the reality conditions, we show that the Type II sigma model at tree-level is consistent with the equations of motion for Type II supergravity in the string gauge.

\end{abstract}
\vskip 0.5cm]

\section{Introduction}

 Low-energy effective actions play an important role in the
study of string theory. Beyond phenomenological applications, they
also provide important pieces of evidence for the existence of
various dual descriptions of string theories.

One way to construct low-energy effective actions in string
theory is looking for the low-energy equations of motion. This is
achieved by defining the sigma-model for the string in a curved background and  requesting conformal invariance \cite{Frad1}.
The low-energy effective action for type II superstring is some N=2
supergravity theory. So, the structure of this supergravity theory
is constrained  by the dynamics of the two-dimensional sigma-model.

 To derive the structure and equations of motion of the N=2
 supergravity theory that represents the Type II low-energy effective
 action, we need to formulate the sigma-model directly in terms of a target superspace and which
  has manifest
local target space supersymmetry. Without such a sigma-model
description, one can only work in components and it is not possible to
determine the off-shell description  of the supergravity selected by
string theory. In addition, it is in general quite difficult to
obtain the fermionic part of the effective action without a
manifestly supersymmetric sigma-model. In the RNS formalism, for
example, the part of the effective action coming from the Ramond
fields is much less understood than the part coming from the
Neveu-Schwarz sector. In the particular case of Type II
superstrings, where there are conjectures relating
non-perturbative states with the Ramond-Ramond  sector \cite{HT},
 this lack of understanding is especially bothersome. On the other hand, in the
 Green-Schwarz formalism, we have manifest SUSY, but the covariant
 quantization is rather impossible due to the kappa symmetry.

A new formalism for the superstring was  discovered by Berkovits
with local N=2 worldsheet superconformal invariance.
 This formalism is known
as hybrid formalism and is related to the RNS formalism by a
field-redefinition \cite{NB1};
it has the advantage of being manifestly spacetime
supersymmetric. It is especially well-suited for compactifications
to four dimensions, where it allows manifestly $4D$
super-Poincar\'e invariant quantization \cite{NB2}.
The coupling of the theory to background fields was discussed in
\cite{NS}, where also the form of the low-energy effective action
was proposed, based on indirect arguments. In \cite{boer}, the low-energy
effective equations of motion of heterotic superstring were
derived directly in superspace by means of this formalism. Although
Berkovits has discovered another formalism based in pure
spinors, which allows manifestly $10D$ super-Poincar\'e covariant
quantization, the hybrid formalism is best understood and it has
been widely used \cite{japas}. In this paper, we adopt this approach
to derive the constraint structure and show consistency of the equations of motion for the
$N=2$ supergravity theory that corresponds to the low-energy
effective theory of Type II superstrings in four dimensions. In the
next section, we describe the Type II sigma-model with local 4D
manifest supersymetry. In Section $3$ we carry out a covariant
background field expansion to check $N=(2,2)$ superconformal
invariance of Type II worldsheet at tree-level. Finally, in Section $4$, we present and discuss the results. The methods presented
here are a generalization of  the methods developed for the heterotic
superstring in \cite{boer}.

\section{Type II sigma-model in the hybrid formalism.}
A  critical N=1 string can be formulated as a critical N=2 string,
without changing the physical content. This is achieved by twisting
the ghost sector of the critical N=1 string \cite{NV}.
 After performing this embedding for the critical
RNS superstring, a field redefinition allows the resulting N=2
string to be made manifestly spacetime supersymmetric for
compactifications to four dimensions. In this case, the critical
$c=6$ matter sector splits into a $c=-3$ four-dimensional part and
a $c=9$ compactification-dependent part. In a flat 4d background,
the type II superstring is in the $N=(2,2)$ superconformal gauge
described by the following action:

\begin{eqnarray}
S &=&\frac{1}{\alpha ^{\prime }}\int d^{2}z\frac{1}{2}\partial x^{m}%
\overline{\partial }x_{m}+p_{\alpha }\overline{\partial }\theta ^{\alpha
}+p_{\dot{\alpha}}\overline{\partial }\theta ^{\dot{\alpha}}++\widehat{p}%
_{\alpha }\partial \widehat{\theta }^{\alpha }\nonumber \\
&+&\widehat{p}_{\dot{\alpha}
}\partial \widehat{\theta }^{\dot{\alpha}}-\frac{\alpha ^{\prime }}{2}%
\overline{\partial }\rho \left( \partial \rho +a_{z}\right)  
+\frac{\alpha ^{\prime }}{2}\partial \widehat{\rho }(\partial \widehat{%
\rho }+\widehat{a}_{z})+S_{c},  \label{acao1}
\end{eqnarray}

where  $S_C$ is the action for the compactification-dependent
superconformal field theory. In this letter, we will not be worried
about the fields that depend on compactificaton, so we need to
concentrate just in the $c=-3$ sector. The four-dimensional
 part of the action contains the spacetime
variables, $x^m$ ($m=0$ to 3), the left-moving fermionic
variables, $\theta^{\a}$ and $\theta^{\dot{\a}}$, 
the conjugate left-moving fermionic variables, $p_{\a}$ and $\bar
{p}_{\dot{\alpha}}$, and one left-moving boson, $\rho$, with a
`wrong' sign for the kinetic term. The right-sector of the Type II
superstring is described by the right-moving fermionic fields,
$\hat{\theta}{}^{\alpha}$,$\hat{\bar{\theta}}{}^{\dot{\alpha}}$,
the conjugate $\hat {p}_{\a}$, $\hat{ \bar p}_{\ad}$, and
one right-moving boson, $\hat{\rho}$. The fields $a_z$, $\hat
{a}_{\bar{z}}$ are the worldsheet $U(1)\times U(1)$ gauge fields
($e^\rho$ carries U(1) charge). The components
$a_{\bar{z}},\hat{a}_z$ can be fixed since the four present gauge fields
fix just two symmetries ($\rho\rightarrow \rho+\mbox{ constant}$ and
$\hat{\rho}\rightarrow \hat{\rho}+\mbox{ constant}$)\footnote{We can see
this by coupling the theory to
 N=(2,2)world-sheet supergravity, which
contains two independent U(1) gauge world-sheet.} The remaining
components, $a_z$ and $\hat{a}_{\bar{z}}$, will be viewed as
Lagrange multipliers imposing the constraints
$\bar{\partial}\rho=0$ and $\partial{\hat{\rho}}=0$, so that $\rho$ and
$\hat{\rho}$ become chiral
 and anti-chiral bosons, respectively. The `wrong' sign for the kinetic terms
of the bosons $\rho$ and $\hat{\rho}$ implies that these fields
 cannot be fermionized, since the OPE's $e^{i\rho(z)}e^{i\rho(w)}=
e^{2i\rho(z)}/(z- -w)$ while $e^{i\rho(z)}e^{-i\rho(w)}= (z -w)$.
It has the same behavior as the negative-energy field, $\phi$,
that appears when bosonizing the RNS ghosts $\gamma=\eta
e^{i\phi}$ and $\beta=\partial\xi e^{-i\phi}$.

The strange $\a '$-dependence of $\rho$ in (\ref{acao1}) will
later be shown to be related to the Fradkin-Tseytlin term.
 Also
this dependence will permit to get the equations for the dilaton
at tree level.
The left-moving $c=-3$ generators  for this N=(2,2) string are:

\begin{eqnarray}
T &=&\left( -\frac{1}{2}\Pi ^{\alpha \dot{\alpha}}\Pi _{\alpha \dot{\alpha}%
}-d_{\alpha }\partial \theta ^{\alpha }-d_{\dot{\alpha}}\partial \overline{%
\theta }^{\dot{\alpha}}+\frac{\alpha ^{\prime }}{2}\partial \rho \partial
\rho +\partial ^{2}\rho \right)\nonumber \\
G &=&\frac{1}{i\alpha ^{\prime }\sqrt{8\alpha ^{\prime }}}\exp \left(
i\rho \right) d^{\alpha }d_{\alpha }, \nonumber \\
 \overline{G}&=&\frac{1}{i\alpha ^{\prime }\sqrt{8\alpha ^{\prime }}}\exp
\left( -i\rho \right) d^{\dot{\alpha}}d_{\dot{\alpha}}  \nonumber \\
J &=&-i\partial \rho \nonumber \\
\label{gera}
\end{eqnarray}
where we have used Pauli matrix to write  vectors in terms of
bi-espinors and we have defined:
\begin{eqnarray}
d_{\alpha } &=&p_{\alpha }+i\overline{\theta }^{\dot{\alpha}}\partial
x_{\alpha \dot{\alpha}}+\frac{1}{2}\overline{\theta }^{2}\partial \theta
_{\alpha }-\frac{1}{4}\theta _{\alpha }\partial \left( \overline{\theta }%
\right) ^{2}  \nonumber \\
d_{\alpha } &=&p_{\dot{\alpha}}+i\overline{\theta }^{\alpha }\partial
x_{\alpha \dot{\alpha}}+\frac{1}{2}\theta ^{2}\partial \overline{\theta }%
_{\alpha }-\frac{1}{4}\overline{\theta }_{\alpha }\partial \left( \overline{%
\theta }\right) ^{2}.  \nonumber \\
\Pi _{a} &\rightarrow &\Pi _{\alpha \dot{\alpha}}=\partial x_{\alpha \dot{%
\alpha}}-\theta _{\alpha }\partial \overline{\theta }_{\dot{\alpha}%
}+i\partial \theta _{\alpha }\overline{\theta }_{\dot{\alpha}}.
\label{def}
\end{eqnarray}


The  right-moving $c=-3$ N=(2,2) generators are


\begin{eqnarray}
\widehat{G} &=&\frac{1}{i\alpha ^{\prime }\sqrt{8\alpha ^{\prime }}}e^{i%
\widehat{\rho }}\widehat{d}^{\alpha }\widehat{d}_{\alpha }
\  \nonumber \\
\widehat{\overline{G}} &=&\frac{1}{i\alpha ^{\prime }\sqrt{8\alpha ^{\prime }%
}}e^{-i\widehat{\rho }}\widehat{d}^{\dot{\alpha}}\widehat{d}_{\dot{\alpha}}
\nonumber \\
\widehat{T} &=&T=\left( -\frac{1}{2}\overline{\Pi }^{\alpha \dot{\alpha}}%
\overline{\Pi }_{\alpha \dot{\alpha}}-\widehat{d}_{\alpha }\overline{%
\partial }\widehat{\theta }^{\alpha }-\widehat{d}_{\dot{\alpha}}\overline{%
\partial }\overline{\widehat{\theta }}^{\dot{\alpha}}+\frac{\alpha ^{\prime }%
}{2}\overline{\partial }\widehat{\rho }\overline{\partial }\widehat{\rho }%
\right)  \label{anti}
\end{eqnarray}
where $\hat{ d}_{\alpha}$ and $\hat{\bar d}_{\dot{\alpha}}$ are
obtained from (\ref{def})  by using hatted variables and
replacing $\partial$ by $\bar{\partial}$. Using the free-field
OPE's the holomorphic (or left-moving) part of the N=(2,2), $c=-3$
algebra can be written as

\begin{eqnarray}
T\left( z\right) T\left( w\right) &=&\frac{c/2}{\left( z-w\right) ^{4}}+%
\frac{2T\left( w\right) }{\left( z-w\right) ^{2}}+\frac{\partial T\left(
w\right) }{\left( z-w\right) },  \nonumber \\
T\left( z\right) G\left( w\right) &=&\frac{\frac{3}{2}G\left( w\right) }{%
\left( z-w\right) ^{2}}+\frac{\partial G\left( w\right) }{\left( z-w\right) }%
,  \nonumber \\
T\left( z\right) \overline{G}\left( w\right) &=&\frac{\frac{3}{2}\overline{G}%
\left( w\right) }{\left( z-w\right) ^{2}}+\frac{\partial \overline{G}\left(
w\right) }{\left( z-w\right) },  \nonumber \\
J\left( z\right) G\left( w\right) &=&\frac{G\left( w\right) }{\left(
z-w\right) },  \nonumber \\
J\left( z\right) \overline{G}\left( w\right) &=&\frac{\overline{G}\left(
w\right) }{\left( z-w\right) },  \nonumber \\
J\left( z\right) J\left( w\right) &=&\frac{c/3}{\left( z-w\right) ^{2}},
\nonumber \\
G\left( z\right) \overline{G}\left( w\right) &=&\frac{\frac{2}{3}c}{\left(
z-w\right) ^{3}}+\frac{2J\left( w\right) }{\left( z-w\right) ^{2}}+\frac{%
2T\left( w\right) +\partial J\left( w\right) }{\left( z-w\right) }.
\label{alge}
\end{eqnarray}


The anti-holomorphic (or right-moving) generators satisfy the same
algebra changing $(z,w)$ for $(\bar z,\bar w)$. The advantage of
the variables $d$ and $\Pi$ over $p$ and $x$ is that they commute
with the target space supersymmetry generators \cite{NS}.

The action also becomes manifestly supersymmetric when expressed in
terms of $d$ and $\Pi$.  This is achieved by  writing the
coordinate of the N=2 flat superspace as  $Z^{A}\rightarrow \left( x^{a},\theta ^{\alpha },\theta ^{%
\dot{\alpha}},\widehat{\theta }^{\alpha },\widehat{\theta }^{\dot{\alpha}%
}\right)$ and defining variables $\Pi^A, \bar{\Pi}^A$ using
vielbeins $E_M{}^A$ to convert curved ($M$) into flat indices
($A$): $\Pi^A=\partial z^M E_M{}^A$ and $\overline{\Pi}^A =
\partial z^M E_M{}^A$, where the indices of the flat and curved N=2
superspace can be written using the $SU(2)$ notation:
$A=(a,\a_{j},\dot{\a}_j)$ and $M=(m,\mu_j,\nu_j)$. The $j= \pm$ is
an $SU(2)$-index and can be raised and lowered using the anti-
symmetric  $\epsilon_{jk}$ tensor. Comparing with the previous
notation
\begin{eqnarray}
\theta ^{\alpha _{+}} &=&\theta ^{\alpha },\qquad \theta ^{\alpha _{-}}=%
\widehat{\theta }^{\alpha }\qquad \overline{\theta }^{\dot{\alpha}_{-}}=%
\overline{\theta }^{\dot{\alpha}}\nonumber \\
 \theta^{\dot{\alpha}_{+}} &=&\widehat{\theta }^{\dot{\alpha}}.
\end{eqnarray}

From the previous section, $\Pi^{a}$ reduces to (\ref{def}) when
$E_M{}^A$ is the vielbein of flat superspace, which is one on the
diagonal and has off-diagonal components:
$E_{\m_{j}}{}^a=\sigma^{a}_{\m\md} \bar{\theta}^{\md}_{j}$.

 The action can now be written directly in the target superspace :
\begin{eqnarray}
S &=&\frac{1}{\alpha ^{\prime }}\int d^{2}z\left[ \frac{1}{2}\Pi ^{\alpha
\dot{\alpha}}\overline{\Pi }_{\alpha \dot{\alpha}}\right. +d_{\alpha }%
\overline{\Pi }^{\alpha _{+}}+d_{\dot{\alpha}}\overline{\Pi }^{\dot{\alpha}%
_{-}}+\widehat{d}_{\alpha }\Pi ^{\alpha _{-}}  \nonumber \\
&&+\widehat{d}_{\dot{\alpha}}\Pi ^{\dot{\alpha}_{+}}+\frac{1}{2}\overline{%
\Pi }^{A}\Pi ^{B}B_{AB}-\frac{\alpha ^{\prime }}{2}\overline{\partial }\rho (%
\overline{\partial }\rho +a_{z})  \nonumber \\
&&\left. \frac{\alpha ^{\prime }}{2}\partial \widehat{\rho }(\partial
\widehat{\rho }+\widehat{a}_{z})\right] \label{mdeloplano},
\end{eqnarray}
where we have introduced an anti-symmetric tensor field $B_{BA}$
whose only non-zero components are:
\begin{eqnarray}
B_{a\alpha _{j}} &=&iC_{\alpha \beta }\theta _{\alpha j}\qquad B_{\alpha
_{+}\beta _{-}}=\theta _{\alpha }\widehat{\theta }_{\beta }  \nonumber \\
B_{\alpha _{+}\dot{\beta}_{+}} &=&\theta _{\alpha }\widehat{\theta }_{\dot{%
\beta}}  \label{vincB}
\end{eqnarray}
 \section{ Sigma-Model in curved N=2 Superspace}

 To formulate the action in a curved background, we can assume that
 the vielbein and the B-field in (\ref{mdeloplano}) have a general form. When expanded to first order around a flat
background, one should recover the massless vertex operators of a
flat background and  the complete set of massless physical states
of  type II superstring should be presented.

 The
massless vertex operators were discussed in \cite{NS}. Due to
manifest supersymmetry, the vertex operators do not distinguish
Ramond and Neveu-Schwarz sectors and may be written in terms of a
superfield that has both sectors. For type II superstring, the
massless compactification-independent vertex operators have the
form:

\begin{equation}
V=\int d^{2}z\{\widehat{\overline{G}},[\widehat{G}\left\{ \overline{G}%
,\left[ G,U\right] \right\} ]\}  \label{verticefinal}
\end{equation}

where $G=\oint \frac{dz}{2\pi i} G(z) = G_{-1/2}$ and similarly
$\hat{G}$ are the $N=(2,2)$ world-sheet supersymmetry generators;
they have precisely the form as one would expect for a theory
with $N=(2,2)$ world-sheet supersymmetry. Furthermore, from the
requirement that the vertex operator produces a state in the BRST
cohomology, it follows that $U$ must be an $N=(2,2)$ primary field of
conformal weight zero and $U(1)\times U(1)$ charge zero, i.e., $U$
should have only single poles in the OPE with
$T,G,\bar{G},\hat{T},\hat{G}, $ and have a regular OPE with $J$
and $\hat J$. These conditions  do not yet completely classify the
inequivalent vertex operators of the theory, because it is
possible to perform certain gauge transformations that do not
change the form of $V$:


\be
 \delta U=\nabla ^{2}\Lambda +\overline{\nabla }^{2}\overline{\Lambda }+
\widehat{\nabla }^{2}\Lambda +\widehat{\overline{\nabla }}^{2}\Lambda
\ee

The N=(2,2) primary field conditions and the gauge conditions
imply that $U$ is a pre-potential for an N=2 conformal supergravity
coupled to a hyper-tensorial multiplet. The gauge fields of
supergravity sit in a Weyl multiplet with 24 bosonic and 24
fermionic off-shell components, while the matter fields are
described by a hyper-tensorial multiplet with  8 bosonic and 8
fermionic off-shell components. This pre-potential represents the
massless compactification-independent fields of the Type II
superstring, without the dilaton. The dilaton does not couple
classically in the action; it is part of the compensator fields
and not part of the hypertensorial multiplet
. To know the precise form of the off-shell N=2
Poincar\`e supergravity that describes the low-energy effective
action for Type II superstrings, we need to know the compensator
and the complete set of supergravity constraints, in particular
the torsion constraints that break the conformal invariance. In a
more conventional formulation, one would use the conformal
invariance to gauge-fix the conformal compensator rather than the
tensor hypermultiplet, but we will show that in the sigma-model it
is the other way around.
 The action in (\ref{mdeloplano}) and the vertex operators provide all
 necessary ingredients to write down the classical part of Type II sigma-model.
 Here, by classical one means zero-order in $\alpha'$. The
 coupling of the dilaton in the sigma-model is described by the
 Fradkin-Tseytlin term, that is, the direct generalization
of the dilaton coupling $\int d^2 z \sqrt{g} R \Phi$ of the
bosonic string 
. So, we need to couple the dilaton
with the N=(2,2) supercurvatures of the worldsheet; this coupling
selects the particular target N=2 superfield that the dilaton will
represent. In this case, worldsheet supersymmetry of the
Fradkin-Tseytlin term requires two different compensators. One
type is described by chiral and anti-chiral
superfields, $\Phi_c$ and $\bar{\Phi}_c$, satisfying
$\nabla_{{\dot\alpha_{-}}}\Phi_c=\nabla_{{\dot\alpha_{-}}}\Phi_c=\nabla_{\alpha_{+}
}\bar{\Phi}_c =\nabla_{\alpha_{-}}\bar\Phi_c=0$. After imposing
the reality condition $(\nabla)^2\Phi_c=
(\hat{\nabla})^2\bar\Phi_c$\footnote{We are using the notation:
$\nabla^2=\nabla^{\a_{+}}\nabla_{\a_{+}}$, $\hat{\nabla}^2=
\nabla^{\a_{-}}\nabla_{\a_{-}}$},
 $\Phi_c$ and $\bar\Phi_c$ can be
identified with the chiral and anti-quiral field strength of a
vector multiplet. This reality condition is not required by
tree-level worldsheet symmetries of the sigma-model,
but it is necessary for
constructing superspace effective actions and can be derived at one loop level.\footnote{In general we
use $\Phi =e^{\phi}$ to construct the low-energy effective action,
while $\phi$ will appear in the sigma-model.}
. The other Type II
compensator is described by twisted-chiral and
twisted-anti-chiral superfields, $\Phi_{tc}$ and $\bar\Phi_{tc}$,
satisfying
$\nabla_{\dot{\alpha_{-}}}\Phi_{tc}=\nabla_{\alpha_{-}}\Phi_{tc}=
\nabla_{\alpha_{+}}\bar\Phi_{tc}=\nabla_{{\dot\alpha}_{+}}\bar\Phi_{tc}=0$.
These fields are related with a tensor hypermultiplet, that is
usually used to fix the $SU(2)$ symmetry of the N=2 conformal
supergravity. Although the twisted-chirality condition on
$\Phi_{tc}$ does not look SU(2) covariant, it can be made
covariant by identifying $\Phi_{tc}$ and $\bar\Phi_{tc}$ with
$L_{--}$ and $L_{++}$ where $L_{jk}$ is the linear field strength
of a tensor hypermultiplet satisfying
$\nabla_{\alpha_{(i}}L_{jk)}=0$. This equation also implies that
$\Phi_{tc}$ satisfies the reality condition
$(\nabla)^2\Phi_{tc}=(\hat{\nabla})^2 \bar\Phi_{tc}$. In the
conformal gauge, the dilaton couples to the $\rho$-field. In this
gauge, the sigma-model is:

\begin{eqnarray}
S &=&\frac{1}{\alpha ^{\prime }}\int d^{2}z\bigg[\frac{1}{2}\Pi ^{\alpha
\dot{\alpha}}\overline{\Pi }_{\alpha \dot{\alpha}}+d_{\alpha }\overline{\Pi }%
^{\alpha _{+}}+d_{\dot{\alpha}}\overline{\Pi }^{\dot{\alpha}_{-}}+\widehat{d}%
_{\alpha }\Pi ^{\alpha _{-}}\nonumber \\
&+&\widehat{d}_{\dot{\alpha}}\Pi ^{\dot{\alpha}%
_{+}} 
+\frac{1}{2}\overline{\Pi }^{A}\Pi ^{B}B_{AB}+d_{\alpha }P^{\alpha \beta }%
\hat{d}_{\beta }+d_{\dot{\alpha}}P^{\dot{\alpha}\dot{\beta}}\hat{d}_{\dot{%
\beta}}\nonumber \\
&+&d_{\alpha }Q^{\alpha \dot{\beta}}\hat{d}_{\dot{\beta}}+d_{\dot{\alpha%
}}\bar{Q}^{\dot{\alpha}\beta }\hat{d}_{\beta }  \nonumber \\
&-&\frac{\alpha ^{\prime }}{2}\left( \overline{\partial }\rho +i\overline{%
\partial }\left( \phi _{c}-\overline{\phi }_{c}+\ \phi _{tc}-\overline{\phi }%
_{tc}\right) \right)\nonumber \\
&\times & \left( \partial \rho +i\partial \left( \phi _{c}-
\overline{\phi }_{c}+\phi _{tc}-\overline{\phi }_{tc}\right) +a_{z}\right)
\nonumber \\
&-&\frac{\alpha ^{\prime }}{2}\left( \partial \widehat{\rho }+i\partial
\left( \phi _{c}-\overline{\phi }_{c}-\ \phi _{tc}+\overline{\phi }
_{tc}\right) \right)\nonumber \\
&\times& \left( \overline{\partial }\widehat{\rho }+i\overline{%
\partial }\left( \phi _{c}-\overline{\phi }_{c}-\ \phi _{tc}+\overline{\phi }%
_{tc}\right) +\widehat{a}_{z}\right) \bigg],\nonumber\\  \label{sigmodel}
\end{eqnarray}
with the previous definitions for the $\Pi$-field, but now with
arbitrary vielbeins and anti-symmetric tensor fields.
$P^{\alpha\dot{\beta}}$ and $Q^{\a\dot{\beta}}$ are chiral and
twisted-chiral field strengths of N=2 conformal supergravity whose
lowest components are the Type II Ramond-Ramond field strengths.
From the $d^\a \hat{d}{}^\a$ and $d^\a \hat{\bar
d}{}^{{\dot\alpha}}$ terms in the supergravity vertex operator,
one sees that at linearized level, $P^{\a\b}=(\Nb)^2 \N^\a
(\hat{\bar{\nabla}})^2 \hat{\nabla}^\b U$ and $Q^{\a\bd}=(\Nb)^2
\N^\a (\hat{\nabla})^2 \nabla{}^\bd U$.
 The sigma-model action contains potentials $B_{BA}$ and
$E_M{}^A$ rather than prepotentials like $U$. Of course, the
anti-symmetric tensor field, the $P^{\alpha \beta}$,
$Q^{\alpha\beta}$ and the vielbeins contain more degrees of
freedom than the prepotential $U$. As usual in supergravity
theories, there are torsion constraints which relate the gauge
fields and field strengths to their prepotentials. In general,
these constraints are imposed by hand; here we will derive these
constraints imposing N=(2,2) superconformal invariance at
tree-level in the sigma-model. 

\section{Perturbative check of the $N=(2,2)$ algebra}

The $N=(2,2)$ algebra derived in (\ref{alge}) for Type II
superstring coupled to flat superspace must be satisfied in the
curved sigma-model. However, in this case we do not have any longer
worldsheet fields satisfying free OPE's and we need a
perturbative approach to check the $N=(2,2)$ algebra. As usual in
string theory, $\alpha'$ counts the number of loops in the two-dimensional
quantum theory. Here, we have an immediate problem
caused by the fields $\rho$ and $\hat{\rho}$. Its kinetic term
does not have an explicit factor of $\frac{1}{\alpha'}$ in front,
and therefore the $\alpha'$-perturbation theory does not make sense
for these fields. In addition, the worldsheet Lagrange multipliers,
$a_z$ and $\hat{a}_{\bar{z}}$, impose the constraints $\bar\partial
(\rho + i(\phi_c-\bar{\phi}_c))=0$ and $\partial (\hat\rho +
i(\phi_tc-\bar{\phi}_tc))=0$, which are difficult to handle. These
last two problems disappear if we make the field-redefinition:

\begin{eqnarray}
\rho  &\rightarrow &\rho -i\left( \phi _{c}-\overline{\phi }_{c}+\ \phi
_{tc}-\overline{\phi }_{tc}\right) ,  \nonumber \\
\widehat{\rho } &\rightarrow &\widehat{\rho }-i\left( \phi _{c}-\overline{%
\phi }_{c}-\ \phi _{tc}+\overline{\phi }_{tc}\right) ;  \label{red}
\end{eqnarray}
after that, $\rho$ and $\hat{\rho}$ become chiral and anti-chiral
bosons, which can be quantized exactly, and which do not interact
with the other fields of the theory. So, after this redefinition,
$\rho$ and $\hat{\rho}$ obey the same free fields OPE's that we
have used to derive the algebra N=2 in (\ref{alge}); for the other
fields, we will use  perturbation theory. Surprisingly, this
redefinition allows to derive information about the dilaton at
tree-level. It is so because the fermionic generators have now the
form:

\begin{eqnarray}
G &\rightarrow &\frac{1}{i\alpha ^{\prime }\sqrt{8\alpha ^{\prime }}}%
\left(e^{i\rho }e^{\phi -\overline{\phi }}d^{\alpha }d_{\alpha }+
\a^{\prime}(...)\right)   \nonumber \\
\overline{G} &\rightarrow &\frac{1}{i\alpha ^{\prime }\sqrt{8\alpha ^{\prime
}}}\left(e^{-i\rho }e^{\overline{\phi }-\phi }d^{\dot{\alpha}}d_{\dot{\alpha}}
+\a^{\prime}(...)\right)\nonumber \\
\widehat{G} &\rightarrow &\frac{1}{i\alpha ^{\prime }\sqrt{8\alpha ^{\prime }%
}}\left(e^{i\widehat{\rho }}e^{\widehat{\phi }-\overline{\widehat{\phi }}}\widehat{d}^{\alpha }\widehat{d}_{\alpha }+\a^{\prime}(...)\right)
\nonumber \\
\widehat{\overline{G}} &\rightarrow &\frac{1}{i\alpha ^{\prime }\sqrt{%
8\alpha ^{\prime }}}\left(e^{-i\widehat{\rho }}e^{\widehat{\overline{\phi }}-%
\widehat{\phi }}\widehat{d}^{\dot{\alpha}}\widehat{d}_{\dot{\alpha}}+\a^{\prime}(...)\right),
  \nonumber \\
\end{eqnarray}


the dots are terms that come from the Fradkin-Tseytlin term and
do not contribute at tree-level. Next, we describe the covariant
background formalism that we intend to perform. A typical beta-function
calculation does not guarantee the full $N=(2,2)$
superconformal invariance. The latter would only follow from a
standard supersymmetric $\beta$-function calculation if the model
could be formulated in $N=(2,2)$ superspace on the worldsheet,
which does not seem possible. So, we need to check the $N=(2,2)$
algebra by calculating the OPE's perturbatively. At tree-level,
there are no double contractions and it is necessary just to
verify  the part of the N=(2,2) algebra that depends on simple
contractions \footnote{We can verify this by checking the orders
in $\alpha'$ in the OPE's (\ref{alge})}. To perform these
calculations, we will use a background covariant expansion that
preserves manifestly all the local symmetries of the target
superspace. In our case, these symmetries are local Lorentz
transformations  and the following $U(1)\times U(1)$
transformations:


\begin{eqnarray}
\delta \phi _{c} &=&\frac{1}{2}\left( \lambda +\widehat{\lambda \text{ }}%
\text{\ }\right) \text{\ \ \ \ \ \ }\delta \overline{\phi }_{c}=-\frac{1}{2}%
\left( \lambda +\widehat{\lambda \text{ }}\text{\ }\right)  \nonumber \\
\delta \phi _{tc} &=&\frac{1}{2}\left( \lambda -\widehat{\lambda \text{ }}%
\text{\ }\right) \text{\ \ \ \ \ \ }\delta \overline{\phi }_{tc}=-\frac{1}{2}%
\left( \lambda -\widehat{\lambda \text{ }}\text{\ }\right)\nonumber \\
\delta \Pi ^{\alpha _{+}} &=&-\frac{1}{2}\lambda \Pi ^{\alpha _{+}}\text{ \
\ }\delta \Pi ^{\dot{\alpha}_{-}}=\frac{1}{2}\lambda \Pi ^{\dot{\alpha}_{-}}%
\text{ }  \nonumber \\
\delta \Pi ^{\alpha _{-}} &=&-\frac{1}{2}\widehat{\lambda }\Pi ^{\alpha _{-}}%
\text{ \ \ }\delta \Pi ^{\dot{\alpha}_{+}}=\frac{1}{2}\lambda \Pi ^{\dot{%
\alpha}_{+}}\text{ }\nonumber \\
\delta d_{\alpha } &=&\frac{1}{2}\lambda d_{\alpha }\text{ \ \ \ \ \ \ \ }%
\delta d_{\dot{\alpha}}=-\frac{1}{2}\lambda d_{\dot{\alpha}}\text{ \ \ \ \ \
\ \ \ }\delta \rho =-2i\lambda \nonumber\\
\delta \widehat{d}_{\alpha } &=&\frac{1}{2}\widehat{\lambda }d_{\alpha }%
\text{ \ \ \ \ \ \ \ }\delta \widehat{d}_{\dot{\alpha}}=-\frac{1}{2}\widehat{%
\lambda }d_{\dot{\alpha}}\text{ \ \ \ \ \ \ \ \ }\delta \widehat{\rho }%
=-2i\lambda
\end{eqnarray}

 The traditional way to achieve such an expansion is to use a
tangent vector that relates two points in target superspace, the
classical field and the quantum fluctuations, then expand all the
tensors in potentials of this vector and use Riemann normal
coordinates to covariantize the expansion.  In this letter, we will
not discuss the details of this expansion 
 and will just put the results. The covariant
derivative in the tangent superspace is:

\begin{eqnarray}
\nabla _{A} &=&E_{A}{}^{M}\partial _{M}+\omega _{A\beta _{+}}{}^{\gamma
_{+}}M_{\gamma _{+}}{}^{\beta _{+}}+\omega _{A\dot{\beta}_{-}}\text{ }^{\dot{%
\gamma}}M_{\dot{\gamma}_{-}}{}^{\dot{\beta}_{-}}  \nonumber \\
&&+\omega _{A\beta _{-}}{}^{\gamma _{-}}M_{\gamma _{-}}{}^{\beta
_{-}}+\omega _{A\dot{\beta}_{+}}\text{ }^{\dot{\gamma}_{+}}M_{\dot{\gamma}%
_{+}}{}^{\dot{\beta}_{+}}+\Gamma _{A}Y+\widehat{\Gamma }_{A}\widehat{Y}\nonumber \\
\label{dcovariant}
\end{eqnarray}
where $\omega,\Gamma, \widehat{\Gamma}$ are the Lorentz and
$U(1)\times U(1)$ connections, $M$ are the Lorentz generators and
$Y,\widehat{Y}$ the $U(1)\times U(1)$ generators. We must observe
that there are two independent spacetime spinors in the Type II
sigma-model, so one has two independent fermionic structure
groups. Thus, just like the two independent U(1) connections one
has two independent sets of irreducible spin connections:
$\omega_{A\a_+}{}^{\b_+}$, $\omega_{A\ad_{-}}{}^{\bd_-}$ and
$\omega_{A\a_{-}}{}^{\b_-}$, $\omega_{A\dot\alpha_{+}}{}^{\bd_+}$.
The covariant derivative satisfies the algebra

\begin{equation}
\left[ \nabla _{C,}\nabla _{A}\right) =T_{CA}{}^{B}\nabla _{B}
+R_{CAE}{}^{D}M_{D}{}^{E}+F_{CA}Y+\widehat{F}_{CA}\widehat{Y}
\end{equation}
where $F$ and $\widehat{F}$ are the super $U(1)\times U(1)$ field
strengths and $T$, $R$ are the supertorsions and supercurvatures.

 For the purpose of our tree-level calculation, we only need to go up to two
background fields in the quadratic part in the quantum fields of
action. With the covariant derivative and the algebra, by using
the algorithm developed in \cite{mukhi,boer}, with the notation:$ A^{\widehat{\a}}=
(A^{\a_-}, A^{\ad_+})$ and $A^{\tilde{\a}}=(A^{\a_+},A^{\ad_-})$,
  we have the action
expanded up to this order :

\begin{eqnarray}
S^{2} &=&\frac{1}{2}\nabla y^{a}\overline{\nabla }y^{a}+d_{\alpha }\overline{%
\nabla }y^{\alpha +}+d_{\dot{\alpha}}\overline{\nabla }y^{\dot{\alpha}-}+%
\hat{d}_{\alpha }\nabla y^{\alpha -}\nonumber \\
&+&\hat{d}_{\dot{\alpha}}\nabla y^{\dot{
\alpha}}  
+\frac{1}{2}\overline{\nabla }y^{a}y^{C}\left( \Pi ^{B}T_{BC}{}^{a}\right) +%
\frac{1}{2}\nabla y^{a}y^{C}\left( \overline{\Pi }^{B}T_{BC}{}^{a}\right)\nonumber \\
&-&\frac{1}{4}\overline{\nabla }y^{C}y^{B}\left( \Pi ^{a}T_{BC}{}^{a}+2\Pi
^{A}H_{ABC}\right)   \nonumber \\
&+&\frac{1}{2}\overline{\nabla }y^{C}y^{B}\left( T_{BC}{}^{\tilde{\alpha}}D_{%
\tilde{\alpha}}\right) +\frac{1}{2}\nabla y^{C}y^{B}\left( T_{BC}{}^{%
\widehat{\alpha }}\hat{D}_{\widehat{\alpha }}\right)\nonumber \\
 &+&d_{\tilde{\alpha}}y^{C}\left( \overline{\Pi }^{B}T_{BC}{}^{\tilde{\alpha}}\right)
+\widehat{d}_{\hat{\alpha}}y^{C}\left( \Pi ^{B}T_{BC}{}^{\hat{\alpha}
}\right)\nonumber \\
&-&\frac{1}{4}\nabla y^{C}y^{B}\left( \overline{\Pi }^{a}T_{BC}{}^{a}
-2\overline{\Pi }^{A}H_{ABC}\right)   \nonumber \\
&+&\frac{1}{4}y^{B}y^{C}\left[ \left( \overline{\Pi }^{D}\overline{\Pi }^{a}-%
\overline{\Pi }^{D}\overline{\Pi }^{a}\right) T_{DCB}{}^{a}-2\overline{\Pi }%
^{D}T_{DCB}{}^{\tilde{\alpha}}D_{\tilde{\alpha}}\right.\nonumber \\
&-&\left. 2\Pi ^{D}T_{DCB}{}^{\widehat{\alpha }}\widehat{D}_{\widehat{\alpha }}\right.
\nonumber \\
&+&\left.
2\overline{\Pi }^{D}\left( \left( -1\right) ^{E\left( D+B\right)
+CD}T_{C}{}^{Ea}T_{DB}{}^{a}+H_{DCB}{}^{E}\right) \Pi _{E}\right]   \nonumber
\\
&+&d_{\alpha }P^{\alpha \beta }\hat{d}_{\beta }+d_{\dot{\alpha}}P^{\dot{
\alpha}\dot{\beta}}\hat{d}_{\dot{\beta}}+d_{\alpha }Q^{\alpha \dot{\beta}}
\hat{d}_{\dot{\beta}}+d_{\dot{\alpha}}\bar{Q}^{\dot{\alpha}\beta }\hat{d}
_{\beta }  \nonumber \\
&+&d_{\alpha }y^{A}\nabla _{A}P^{\alpha \beta }\widehat{D}_{\beta }+d_{\dot{
\alpha}}y^{A}\nabla _{A}P^{\dot{\alpha}\dot{\beta}}\widehat{D}_{\dot{\beta}
}\nonumber \\
&+&d_{\alpha }y^{A}\nabla _{A}Q^{\alpha \dot{\beta}}\widehat{D}_{\dot{\beta}
}+d_{\dot{\alpha}}y^{A}\nabla _{A}\bar{Q}^{\dot{\alpha}\beta }\widehat{D}
_{\beta }  \nonumber \\
&+&D_{\alpha }y^{A}\nabla _{A}P^{\alpha \beta }\hat{d}_{\beta }+D_{\dot{\alpha%
}}y^{A}\nabla _{A}P^{\dot{\alpha}\dot{\beta}}\hat{d}_{\dot{\beta}}+D_{\alpha
}y^{A}\nabla _{A}Q^{\alpha \dot{\beta}}\hat{d}_{\dot{\beta}}\nonumber \\
&+&D_{\dot{\alpha}}y^{A}\nabla _{A}\bar{Q}^{\dot{\alpha}\beta }\hat{d}_{\beta }  +D_{\alpha }y^{A}y^{B}\nabla _{B}\nabla _{A}P^{\alpha \beta }\widehat{D}
_{\beta }\nonumber\\
&+&D_{\dot{\alpha}}y^{A}y^{B}\nabla _{B}\nabla _{A}P^{\dot{\alpha}
\dot{\beta}}\widehat{D}_{\dot{\beta}}
+D_{\alpha }y^{A}y^{B}\nabla _{B}\nabla
_{A}Q^{\alpha \dot{\beta}}\widehat{D}_{\dot{\beta}}  \nonumber \\
&+&D_{\dot{\alpha}}y^{A}y^{B}\nabla _{B}\nabla _{A}\bar{Q}^{\dot{\alpha}%
\beta }\widehat{D}_{\beta },  \label{s2}
\end{eqnarray}

where:
\begin{eqnarray}
T_{DCB}{}^{A}&=&R_{DCB}{}^{A}+\omega \left( A\right) F_{DC}\delta _{B}{}^{A}+%
\widehat{\omega }\left( A\right) \widehat{F}_{DC}\delta
_{B}{}^{A}\nonumber \\
&+&T_{DC}{}^{E}T_{EB}{}^{A}+\left( -1\right) ^{CD}\nabla
_{C}T_{DB}{}^{A}  \label{tgrande}\nonumber\\
H_{DCB}{}^{A}&=&\nabla _{C}H_{DB}{}_{A}\left( -1\right)
^{CD}\nonumber \\
&-&T_{CA}{}^{E}H_{EDB}{}\left( -1\right) ^{A\left( B+D\right)
+CD}
+T_{DC}{}^{E}H_{EBA} \nonumber 
\end{eqnarray}


$\omega(A)$ and $\hat{\omega}(A)$ are the $U(1)\times U(1)$
weights of $A$. The only ones different from zero are:
$\omega(\alpha_{+})= \hat{\omega}(\alpha_{+})
=\frac{1}{2}$,$\omega(\dot\alpha_{-})=
\hat{\omega}(\dot\alpha_{+})= -\frac{1}{2}$. In the expanded
action, we have the quantum fields: $Y^{A}$, $d_\a$,$\hat d_\alpha$
and the background fields: $\Pi^{A}$, $D_\alpha$, $\hat D_\alpha$
. Since the $d_\alpha$ and $\hat d_\alpha$ fields are space-time
independent, we have chosen a simple expansion for these
 fields, that preserves the superspace symmetries:
 $d_{\a}=d_{\a}+D_{\a}$ and similarly for $\hat{d}_{\a}$.

Now, we can describe the kind of calculation we intend to do. The
kinetic part of the action provides the worldsheet propagators:
\begin{equation}
\langle d_{\alpha}y^{\beta_{+}}\rangle= \frac{\d_{\a}{}^{\b} }{z-w}
\qquad
 \langle \hat{d}_{\alpha}y^{\beta_{-}}\rangle=\frac{\d_{\a}{}^{\b}}{\overline{z}-\overline{w}}
\end{equation}
and  the same for the dot spinors. The bosonic fields have the
two dimensional propagator:
$\langle{y^{a}y^{b}}\rangle=\eta^{ab}\ln|z-w|$. The other part of
the action provides the vertices. By expanding the generators
using the same background covariant  expansion we can calculate
the tree-level OPE's contracting the fields with the vertices in
the action. By demanding that the N=(2,2) algebra be satisfied, we get
the torsion constraints  of N=2 supergravity. All the results we
will get from the fermionic part of the algebra.


Besides Lorentz and $U(1)\times U(1)$ invariance, the background
field expansion of the action has an additional set of symmetries,
which we denote by `shift symmetries'. These originate from the fact
that the original action depends only on the vielbeins, not on
torsions and curvatures. In our case the shift symmetry has the
form:

\begin{eqnarray}
\delta \omega _{AB}{}^{C} &=&Y_{AB}{}^{C}  \nonumber \\
\delta \Gamma _{A} &=&X_{A}  \nonumber \\
\delta \widehat{\Gamma }_{A} &=&\widehat{X}_{A}  \nonumber \\
\delta T_{AB}{}^{C} &=&Y_{\left[ AB\right\} }{}^{C}+\omega \left( C\right)
X_{\left[ A\right. }\delta _{\left. B\right) }{}^{C}+\widehat{\omega }\left(
C\right) \widehat{X}_{\left[ A\right. }\delta _{\left. B\right) }{}^{C}
\nonumber \\
\delta y^{A} &=&\frac{1}{2}y^{B}y^{C}\left( Y_{CB}{}^{A}++\omega \left(
A\right) X_{C}\delta _{B}{}^{A}+\widehat{\omega }\left( A\right) \widehat{X}%
_{C}\delta _{B}{}^{A}\right) 
  \nonumber \\
\delta d_{\alpha } &=&\left( y^{M}Y_{M\alpha }{}^{\beta }+\frac{1}{2}%
y^{A}X_{A}\delta _{\alpha }{}^{\beta }\right) \left( d_{\beta }+D_{\beta
}\right)
   \nonumber \\
\delta \widehat{d}_{\alpha } &=&\left( y^{M}Y_{M\alpha }{}^{\beta }+\frac{1}{%
2}y^{A}\widehat{X}_{A}\delta _{\alpha }{}^{\beta }\right) \left( \widehat{d}%
_{\beta }+\widehat{D}_{\beta }\right)
 .  \label{shift}
\end{eqnarray}


This symmetry must be manifest in the OPE's and we will use it to fix
some components of the torsions, providing the  `conventional
constraints'.

We are now ready to check the superconformal invariance  at tree-level.
The generators must satisfy the tree-level part of the
OPE's (simple contractions) and in addition the left-moving
generators should be holomorphic and the right-moving
anti-holomorphic. By requesting the conditions $\overline{\partial}
G=\partial\widehat{G}=0$ with help of the equations of motion,
\begin{eqnarray}
\bar{\Pi}^{\alpha +}+P^{\alpha \beta }\widehat{d}_{\beta }+Q^{\alpha \dot{%
\beta}}\widehat{d}_{\dot{\beta}} &=&0  \nonumber \\
\bar{\Pi}^{\dot{\alpha}-}+P^{\dot{\alpha}\dot{\beta}}\widehat{d}_{\dot{\beta}%
}+\overline{Q}^{\dot{\alpha}\beta }\widehat{d}_{\beta } &=&0  \nonumber \\
\Pi ^{\alpha -}-d_{\beta }P^{\beta \alpha }-d_{\dot{\beta}}\overline{Q}^{%
\dot{\beta}\alpha } &=&0  \nonumber \\
\Pi ^{\dot{\alpha}+}-d_{\dot{\beta}}P^{\dot{\beta}\dot{\alpha}}-d_{\beta
}Q^{\beta \dot{\alpha}} &=&0 \nonumber 
\end{eqnarray}

\begin{eqnarray}
\overline{\nabla }d_{\tilde{\alpha}}&+&\frac{1}{2}\left( \Pi ^{C}T_{C\tilde{%
\alpha}}{}^{b}\overline{\Pi }_{b}+\overline{\Pi }^{C}T_{C\tilde{\alpha}%
}{}^{b}\Pi _{b}\right) -\overline{\Pi }^{C}\left( T_{C\tilde{\alpha}}{}^{%
\tilde{\beta}}d_{\tilde{\beta}}\right)\nonumber \\
&-&\Pi ^{C}\left( T_{C\tilde{\alpha}
}{}^{\hat{\beta}}\widehat{d}_{\hat{\beta}}\right) -
\Pi ^{C}\overline{{\Pi }}^{B}H_{BC\tilde{\alpha}}\nonumber \\
&+&d_{\beta }\left( \nabla _{
\tilde{\alpha}}P^{\beta \gamma }\widehat{d}_{\gamma }+\nabla _{\tilde{\alpha}%
}Q^{\beta \dot{\gamma}}\widehat{d}_{\dot{\gamma}}\right)
\nonumber \\
&+&d_{\dot{\beta}}\left( \nabla _{\tilde{\alpha}}P^{\dot{\beta}\dot{\gamma}}%
\widehat{d}_{\dot{\gamma}}+\nabla _{\tilde{\alpha}}\bar{Q}^{\dot{\beta}%
\gamma }\widehat{d}_{\gamma }\right) =0
  \label{eqmov}
\end{eqnarray}

we get

\begin{eqnarray}
T_{\tilde{\b}\widehat{\a}c}&=&2H_{\tilde{\b}\widehat{\a}c}= 0\nonumber \\
T_{\dot{\b}_{-}\a_{+}c}&-&2H_{c\dot{\beta}_{-}\alpha _{+}}= 0 \nonumber \\
T_{\dot{\b}_{+}\a_{-}c}&+& 2H_{c\dot{\beta}_{+}\alpha _{-}}= 0 \nonumber \\
T_{\dot{\b}_{-}\dot{\a}_{-}c}&-&2H_{c\dot{\b}_{-}\dot{\a}_{-}}=
T_{{\b}_{-}\a _{-}c}-2H_{c\b_{-}\a_{-}}=0 \nonumber\\
T_{\dot{\beta}_{+}\dot{\a} _{+}c}&-&2H_{\dot{\beta}_{+}c\dot{\a} _{+}}=
T_{{\b}_{+}\a _{+}c}-2H_{c\b_{+}\a_{+}}=0 \nonumber\\
T_{a\tilde{\beta}c}&+&T_{ac\tilde{\beta}}=H_{a\tilde{\beta}c}=0.
\nonumber\\
T_{a\hat{\beta}c}&+&T_{ac\hat{\beta}}=H_{a\hat{\beta}c}=0.
\nonumber\\
T_{c\alpha _{+}}{}^{\gamma _{-}} &=&-T_{\alpha _{+}\dot{\alpha}_{-}c}%
\overline{Q}^{\dot{\alpha}\gamma }, \qquad
T_{c\alpha _{+}}{}^{\dot{\gamma}_{+}} =T_{\alpha _{+}\dot{\alpha}_{-}c}P^{%
\dot{\alpha}\dot{\gamma}},\nonumber \\ 
T_{c\alpha +}{}^{\dot{\gamma}_{-}} &=&T_{c\alpha -}{}^{\dot{\gamma}_{+}}=0
\end{eqnarray}
In addition to these constraints, we have the equations of motion for the dilaton:

\begin{eqnarray}
\nabla _{\beta }P^{\beta \gamma} +P^{\beta \gamma}T_{\beta
\alpha }{}^{\alpha }&+&Q^{\dot{\beta}\gamma }T_{\dot{\beta}\alpha }{}^{\alpha
}+P^{\alpha \beta _{-}}T_{\alpha \beta _{-}}{}^{\gamma _{-}}  \nonumber \\
=P^{\beta \gamma _{-}}\nabla _{\beta }\left( \phi -\bar{\phi}\right) &+&Q^{%
\dot{\beta}\gamma }\nabla _{\dot{\beta}}\left( \phi -\bar{\phi}\right)\nonumber\\
\nabla _{\beta _{-}}P^{\gamma \beta _{-}} -P^{\gamma \beta _{-}}T_{\beta
_{-}\alpha _{-}}{}^{\alpha _{-}}&-&Q^{\gamma \dot{\beta}_{+}}T_{\dot{\beta}%
_{+}\alpha _{-}}{}^{\alpha _{-}}-P^{\alpha \beta _{-}}T_{\alpha \beta
_{-}}{}^{\gamma }  \nonumber \\
=-P^{\gamma \beta }\nabla _{\beta _{-}}\left( \widehat{\phi }-\widehat{%
\bar{\phi}}\right)&-&Q^{\gamma \dot{\beta}_{+}}\nabla _{\dot{\beta}%
_{+}}\left( \widehat{\phi }-\widehat{\bar{\phi}}\right)\nonumber 
\end{eqnarray}

\begin{eqnarray}
T_{\gamma \alpha _{-}}{}^{\alpha _{-}}+\nabla _{\gamma }\left( \widehat{\phi
}-\widehat{\bar{\phi}}\right) &=&0 \quad
T_{\gamma \dot{\alpha}_{+}}{}^{\dot{\alpha}_{+}}-\nabla _{\gamma }\left(
\widehat{\phi }-\widehat{\bar{\phi}}\right) =0
  \nonumber \\
T_{a\alpha _{-}}{}^{\alpha _{-}}+\nabla _{a}\left( \widehat{\phi }-\widehat{%
\bar{\phi}}\right) &=&0 \quad
 T_{a\dot{\alpha}_{+}}{}^{\dot{\alpha}_{+}}-\nabla _{a}\left( \widehat{\phi }-
\widehat{\bar{\phi}}\right) =0 \nonumber\\
T_{\gamma _{-}\alpha }{}^{\alpha }+\nabla _{\gamma _{-}}\left( \phi -\bar{%
\phi}\right) &=&0 \quad
T_{\gamma _{-}\dot{\alpha}}{}^{\dot{\alpha}}-\nabla _{\gamma _{-}}\left(
\phi -\bar{\phi}\right) =0
 \nonumber \\
T_{a\alpha }{}^{\alpha }+\nabla _{a}\left( \phi -\bar{\phi}\right) &=&0
\quad T_{a\dot{\alpha}}{}^{\dot{\alpha}}-\nabla _{a}\left( \phi -\bar{\phi}\right)
=0 \label{cn}
\end{eqnarray}



The tree-level  OPE between the holomorphic generators
 and anti-holomorphic ones must be regular. Using the background expansion for the generators, we can write the $G(z)\widehat{G}(w)$ OPE as
\begin{eqnarray}
G(z)\widehat{G}(w)&=&D_{\alpha }\left( z\right) \left\langle d^{\alpha }\left( z\right) \widehat{d
}^{\beta }\left( w\right) \right\rangle \widehat{D}_{\beta }\left( w\right)
\nonumber \\
&+&D_{\alpha }\left( z\right) \left\langle d^{\alpha }\left( z\right)
y^{A}\left( w\right) \right\rangle \widehat{D}^{2}\left( w\right) \nabla
_{A}\left( \widehat{\phi }-\widehat{\overline{\phi }}\right) \nonumber \\   
&+&D^{2}\left( z\right)\nabla
_{A}\left( \phi -\overline{\phi }\right) \left\langle y^{A}\left( z\right) \widehat{d}^{\alpha
}\left( z\right) \left( w\right) \right\rangle \widehat{D}_{\alpha }
\nonumber \\  
&+&D^{2}\left( z\right) \left\langle y^{A}\left( z\right) y^{B}\left( w\right)
\right\rangle \widehat{D}^{2}\left( w\right) \nabla _{A}\left( \phi -%
\overline{\phi }\right)\nonumber \\
&\times& \nabla _{B}\left( \widehat{\phi }-\widehat{\overline{%
\phi }}\right).  \label{yy}
\end{eqnarray}

In this expression we did not write the exponential terms involving the
$\rho$ and $\phi$ fields. By contracting the quantum fields in the
expectation values with vertices that came from the action
($\ref{s2}$), we have many non-regular terms with 3 and 4
background fields. We can show that, when we use the previous
constraints and equations of motion, all the non-regular terms are
cancelled,
 as we should expect. However, the tree-level part of
the $\langle G(z)G(w)\rangle$ is

\begin{equation}
\left\langle G\left( z\right)G\left( w\right) \right\rangle =
\frac{1}{\left( z-w\right)}\left\langle... \right\rangle
\end{equation}
where we have used the $e^{i\rho(z)}e^{i\rho(w)}$ OPE and the dots
are terms similar to (\ref{yy}), with the hatted fields replaced
by non-hatted fields.  Here, we get new information. To get the
right N=(2,2) algebra, the expectation values must be of order
$O(z-w)$. Again, we contract the quantum fields with the vertices
of the action and the  anti-holomorphic part of these OPE's are
cancelled by the conditions (\ref{cn}). From the holomorphic part, we
get a regular term in the expectation values that give us the
constraints
\begin{eqnarray}
T_{\gamma _{+}\alpha _{+}}{}^{a}&=&T_{\gamma _{+}\alpha _{+}}{}^{\dot{\gamma}%
_{i}}=H_{\gamma _{+}\alpha _{+}a}=H_{\gamma _{+}\alpha _{+}\dot{\gamma}%
_{i}}=H_{\gamma _{+}\alpha _{+}\gamma _{i}}=0 \nonumber \\ 
T_{\gamma _{+}\alpha _{+}}{}^{\alpha _{+}}&+&2\nabla _{\gamma }\left( \phi -%
\overline{\phi }\right) =0  \label{vin2}
\end{eqnarray}

This new information comes from the fact that we have exact
conformal theory for the $\rho$ fields. The results for $\langle
G(z)G(w)\rangle$ are similar, changing $\gamma_{+}$,$ \a_{+}$ for $\gamma_{-}$,$ \a_{-}$.

The OPE's must be shift-invariant. This is true for all the
derived constraints and equations of motion, except for the last
equation in (\ref{vin2}). This is another consequence of the $\rho$-field.
Fortunately, there is a systematic way to cancel this equation.
We can redefine the expansion for the $d_\alpha$-field, showing
that the shift invariance fixes the expansion for this field or add
the zero-term to the action:

\[
\int d^{2}z\left( \overline{\nabla }D_{\gamma }+...\right) y^{\alpha
_{+}}y^{\beta _{+}}S_{\beta _{+}\alpha _{+}}{}^{\gamma _{+}},
\]
where $(\overline{\nabla} D_{\gamma} + \ldots)$ is the field
equation  with $d$ replaced by $D$. Since $D$ is on-shell , the
extra term is always zero. If one partially integrates it, one gets a
series of vertices with two background fields, that does not give
any contribution  and precisely one vertex with one background
field, namely :$\int d^{2}z\left( -2D_{\gamma }\overline{\partial
}y^{\alpha _{+}}y^{\beta _{+}}S_{\beta _{+}\alpha _{+}}{}^{\gamma
_{+}}\right)$. We can choose $S_{\alpha\b}{}^{\b}$ to cancel the
equation and as it may be verified, this term in the action does not
give any new contribution; the same trick was used in \cite{boer}
and \cite{banks}.

 Before writing down the
constraints and equation of motion, we can check that  no
information came from the OPE's $\langle
G(z)\overline{G}(w)\rangle$ and
$\langle\hat{G}(z)\hat{\overline{G}}(w)\rangle$. While the tree-level OPE's
$\langle d^2(z)d^2(w)\rangle$ and
$\langle\hat{d}^2(z)\hat{d^2}(w)\rangle$ must be of order ${\cal
O}((z-w))$ and ${\cal O}((\bar{z}-\bar{w}))$, the OPE's $\langle
d^2(z)\bar{d}^2(w)\rangle$ and
$\langle\hat{d}^2(z)\hat{\bar{d}^2}(w)\rangle$ must be of order
${\cal O}((z-w)^{-1})$ and ${\cal O}((\bar{z}-\bar{w})^{-1})$. Since
terms with more than two backgrounds fields are always of
order ${\cal O}((z-w)^{-1})$, they are not relevant and we have
just terms with two background fields: $\langle
d^2(z)\bar{d}^2(w)\rangle$ and
$\langle\hat{d}^2(z)\hat{\bar{d}^2}(w)\rangle$, that are trivially
zero since we do not have propagators from $d_{\a}$ to $\bar{
d}_{\dot \alpha}$ and from $\hat{d}_{\a}$ to $\hat{\bar d}_{\dot
\alpha}$. The OPE's involving $T$ and  $\hat {T}$ are satisfied by
using the equations of motion (\ref{eqmov}) and identities like
$\overline{\nabla}\Pi^{\alpha_i}-\nabla\overline{\Pi}^{\alpha_i}
=\Pi^{A}\overline{\Pi}^{B}T_{AB}{}^{\alpha_i}$. We will discuss
first the constraints; they are divided into the following
categories:

A. Representation-preserving constraints:
\begin{eqnarray}
T_{\alpha _{k}\dot{\beta}_{+}}{}^{\dot{\gamma}_{-}} &=&T_{\alpha
_{j}\beta
_{k}}{}^{c}=0  \nonumber \\
T_{\alpha _{k}\beta _{+}}{}^{\dot{\gamma}_{-}} &=&T_{\alpha _{k}\dot{\beta}%
_{-}}{}^{\dot{\gamma}_{+}}=0
\end{eqnarray}

B. Conformal Constraints:
\begin{eqnarray}
H_{\alpha _{i}\dot{\beta}_{j}}{}^{c} &=&-i\delta _{ij}\delta
_{\alpha
}^{\lambda }\delta _{\dot{\beta}}^{\dot{\lambda}}  \nonumber \\
T_{c\alpha _{+}}{}^{\gamma _{-}} &=&-T_{\alpha _{+}\dot{\alpha}_{-}c}%
\overline{Q}^{\dot{\alpha}\gamma },T_{c\alpha _{-}}{}^{\gamma
_{+}}=T_{\alpha _{-}\dot{\alpha}_{+}c}Q^{\gamma \dot{\alpha}}  \nonumber \\
T_{c\alpha _{+}}{}^{\dot{\gamma}_{+}} &=&T_{\alpha _{+}\dot{\alpha}_{-}c}P^{%
\dot{\alpha}\dot{\gamma}},\text{ }T_{c\alpha _{-}}{}^{\dot{\gamma}%
_{-}}=T_{\alpha _{-}\dot{\alpha}_{+}c}P^{\dot{\gamma}\dot{\alpha}}
\nonumber
\\
T_{c\alpha +}{}^{\dot{\gamma}_{-}} &=&T_{c\alpha
-}{}^{\dot{\gamma}+}{}=0
\end{eqnarray}

The first type of constraints is the usual representation-
preserving constraints which allow a consistent definition of
chiral and twisted-chiral superfields. The second type of
constraints is conformal-breaking constraints, which are
necessary since the sigma-model action is not invariant under the
spacetime scale transformations that transform $\d E_a{}^M=\L
E_a{}^M$.  The relation between the $P^{\alpha\beta}$ and
$Q^{\alpha\dot{\beta}}$ with the torsions shows us a clear
geometrical meaning for the Ramond-Ramond fields in the
superspace.  In general, we can write the hypertensorial multiplet
$H_{ABC}$ in terms of a linear multiplet ,
$H_{\alpha_{i}\dot{\beta}_{j}}{}^a= \sigma^{
a}_{\alpha\dot{\beta}}L_{ij}$, where the lowest component of the
linear multiplet is a $SU(2)$ triplet $l_{ij}$. So, also the
$SU(2)/U(1)$ invariance is broken in the Type II sigma-model by
gauge-fixing $L_{jk}=\delta_{jk}$. In addition, the Bianchi
identities imply that $T_{abc}=-2H_{abc}$. So, the Type II
superstring selects a particular gauge, where the matter is fixed and
breaks the scale and  $SU(2)/U(1)$ symmetries of the conformal
supergravity. Part of the hypermultiplet that is not fixed goes to
the supergravity multiplet which, after imposing  the conventional
constraints, presents
 $32+32$ off-shell degrees of freedom. In general, when passing from
conformal supergravity to Poincar\`e supergravity, the compensator
fields are fixed. For a historical view of the development of the
$N=2$, $4d$ sugra in supersace see ref \cite{gates}. In particular,
the third reference discusses for the first time the change of the
 constraints that could lead to the string frame formulation derived here. In the last reference the vector and tensorial compensators are discussed \footnote{I would like to thank Professor Jim Gates for clarifying this point to me}. In \cite{wit}, a ``minimal multiplet"
is employed as a starting point, consisting of N= 2 conformal
supergravity coupled to a vector multiplet, and by coupling this
minimal multiplet to chiral, a non-linear, or a hypertensorial
multiplet we have 3 different off-shell Poincar\`e supergravity.
In particular, in the last one, the vector multiplet fixes the
scale and $U(1)$ invariance while the hypermultiplet fixes
$SU(2)/U(1)$.
Here, we show that the Type II supergravity is a $U(1)\times U(1)$
version of this N=2 Poincar\'e supergravity, when the matter is
fixed and the compensators are dynamical.
 Let us now determine a maximal set of conventional constraints. From our sigma-model
 point of view, part of the conventional constraints can be derived from the OPE's and
 the other part can  be viewed as a gauge fixing of
the shift symmetry discussed previously, choosing $X_A$,
$Y_{AB}{}^C$, $\hat{X}_A$ and $\hat{Y}_{AB}{}^C$ properly.

C. Conventional constraints:
\begin{eqnarray} T_{\alpha _{i}\dot{\beta}_{j}}{}^{c}
&=&-2i\varepsilon _{ij}\delta _{\alpha
}^{\lambda }\delta _{\dot{\beta}}^{\dot{\lambda}}  \nonumber \\
T_{\alpha _{i}(bc)} &=&0. \nonumber \\
T_{\alpha _{+}\beta _{+}}^{\text{ \ \ \ \ \ }\gamma _{+}} &=&T_{\dot{\alpha}%
_{i}\beta _{+}}^{\text{ \ \ \ \ \ }\gamma _{+}}=T_{\alpha _{i}\dot{\beta}%
_{-}}^{\text{ \ \ \ \ }\dot{\gamma}_{-}}=T_{a(\beta _{+}}^{\text{ \ \ \ \ \ }%
\gamma _{+})}=T_{a(\dot{\beta}_{+}}^{\text{ \ \ \ \ \ }\dot{\gamma}%
_{+})}=0\nonumber \\
T_{a\beta _{+}}^{\text{ \ \ \ \ }\beta _{+}}&-&T_{a\dot{\beta}_{-}}^{%
\text{ \ \ \ }\dot{\beta}_{-}}=0  \nonumber \\
T_{\alpha _{-}\beta _{-}}^{\text{ \ \ \ \ \ }\gamma _{-}} &=&T_{\dot{\alpha}%
_{i}\beta _{-}}^{\text{ \ \ \ \ }\gamma _{-}}=T_{\alpha _{i}\dot{\beta}%
_{+}}^{\text{ \ \ \ \ }\dot{\gamma}_{+}}=T_{a(\beta _{-}}^{\text{ \ \ \ \ \ }%
\gamma _{-})}=T_{a(\dot{\beta}_{+}}^{\text{ \ \ \ \ }\dot{\gamma}%
_{+})}=0 \nonumber \\
T_{a\beta _{-}}^{\text{ \ \ \ \ }\beta _{-}}&-&T_{a\dot{\beta}_{+}}^{%
\text{ \ \ \ }\dot{\beta}_{+}}=0
\end{eqnarray}
These constraints define the vector components of the
super-vielbein in terms of the spinor components, and the spin
connections in terms of the super-vielbein. The equations of
motion for the compensators become

\begin{eqnarray}
\nabla _{\widehat{\gamma }}\left( \phi _{c}-\overline{\phi }_{c}+\
\phi
_{tc}-\overline{\phi }_{tc}\right) &=&0  \nonumber \\
\nabla _{a}\left( \phi _{c}-\overline{\phi }_{c}+\ \phi
_{tc}-\overline{\phi
}_{tc}\right) &=&0  \nonumber \\
\nabla _{\gamma }\left( \phi _{c}-\overline{\phi }_{c}-\ \phi _{tc}+%
\overline{\phi }_{tc}\right) &=&0  \nonumber \\
\nabla _{a}\left( \phi _{c}-\overline{\phi }_{c}-\ \phi
_{tc}+\overline{\phi }_{tc}\right) &=&0.  \label{eqarvore}
\end{eqnarray}

With the help of Bianchi identities the equation involving $\nabla
_{\beta }P^{\beta \gamma}$  can be written as
\be
 \left[ \nabla _{a},\nabla _{%
\dot{\gamma}_{+}}\right] \left( \phi _{c}-\overline{\phi }_{c}+\ \phi _{tc}-%
\overline{\phi }_{tc}\right)=0 \nonumber ,
\ee
 so it is cancelled by
(\ref{eqarvore}).These equations of motion are precisely the equations of motion that describe the 16+16 degrees of freedom of the N=2 compensators coupled to N=2 supergravity in the string gauge. At this point we need to emphasize that this is true if the dilaton
fields satisfy the reality conditions discussed previously, which
are not derived at tree level. So, strictly speaking, we just showed that the equations of motion obtained here are consistent with the equations of motion for Type II supergravity in the string gauge. To these equations make sense, we need to impose the reality conditions, that appear in one loop level \cite{tese}. These results shall soon be reported elsewhere. The corrections discussed in \cite{portugues} will appear in higher loops.

It is unusual to get any information about the dilaton at tree
level. This is other immediate consequence of the $\rho$ field
behavior. We can see that the $\rho$ field has no $\a'$-dependence
in the action, but appears with the same order of $\a'$ as the $d$
fields in the fermionic generators in (\ref{gera}); this is a
particularity of this formalism.
 In addition, to have a consistent
perturbation theory in the superconformal gauge we need to make
the redefinition (\ref{red}). So, we have a new dependence of the
tree level part of the generators on $e^{\phi_c}$ and
$e^{\phi_{tc}}$. This generates the conditions (\ref{eqarvore}) but
not the reality conditions, which appear at one loop.

The  low-energy effective action for Type II superstring in four
dimensions must reproduce these equations of motion. In the work
of ref. \cite{NS}, by using indirect arguments and not by checking
directly the superconformal invariance, a Type II low-energy
effective action was proposed in the harmonic superspace. We can
check that this action generates the equations of motion showed
here.

\section*{Acknowledgments:} I would like to thank Nathan Berkovits for suggesting this problem and for useful discussions, and especially Jos\'e Abdalla Helayel-Neto and D\'afni Marchioro for useful discussions, suggestions and motivation. This research is part of the author PhD thesis, advised by Nathan Berkovits and supported by FAPESP grant 98/06240-1.

\end{document}